\documentclass[twocolumn,showpacs,preprintnumbers,amssymb]{revtex4}
\usepackage{graphicx}
\usepackage{dcolumn}
\usepackage{bm}
\graphicspath{{./fig/}{./png/}}
\usepackage {eepic}
\usepackage {rotating}

\newcommand{\ovec}[1]{{\mbox{\boldmath $#1$}}}

\newcommand{\bb}{\ovec{b}}
\newcommand{\bB}{\ovec{B}}

\newcommand{\be}{\ovec{e}}
\newcommand{\bscE}{\ovec{\cal{E}}}
\newcommand{\cE}{\cal{E}}

\newcommand{\bg}{\ovec{g}}
\newcommand{\bh}{\ovec{h}}
\newcommand{\bJ}{\ovec{J}}

\newcommand{\bu}{\ovec{u}}

\newcommand{\bx}{\ovec{x}}

\newcommand{\bzo}{{\bf 0}}

\newcommand{\bmB}{\overline{\ovec{B}}}
\newcommand{\mB}{\overline{B}}

\newcommand{\bpsi}{\ovec{\psi}}
\newcommand{\bxi}{\ovec{\xi}}
\newcommand{\bomega}{\ovec{\omega}}
\newcommand{\bOmega}{\ovec{\Omega}}
\newcommand{\p}{\partial}
\newcommand{\bnab}{\ovec{\nabla}}
\newcommand{\x}{\times}
\def\ol{\overline}
\def\iu {\mbox{i}}
\def\dd {\mbox{d}}

\begin{document}
\preprint{NORDITA 2008-1}

\title{Alpha--effect dynamos with zero kinetic helicity}
\author{Karl-Heinz R\"adler}
\affiliation{Astrophysical Institute Potsdam,
An der Sternwarte 16, D-14482 Potsdam, Germany}
\author{Axel Brandenburg}
\affiliation{NORDITA, Roslagstullsbacken 23, SE-10691 Stockholm, Sweden}
\date{$ $Revision: 1.28 $ $}

\begin{abstract}
A simple explicit example of a Roberts--type dynamo is given in which the $\alpha$--effect
of mean--field electrodynamics exists in spite of point--wise vanishing kinetic helicity of the fluid flow.
In this way it is shown that $\alpha$--effect dynamos do not necessarily require non--zero kinetic helicity.
A mean--field theory of Roberts--type dynamos is established within the framework of
the second--order correlation approximation.
In addition numerical solutions of the original dynamo equations are given,
that are independent of any approximation of that kind.
Both theory and numerical results demonstrate the possibility of dynamo action
in the absence of kinetic helicity.

\end{abstract}


\maketitle

\section{Introduction}
\label{intro}

The essential breakthrough in the understanding of the origin of the large--scale magnetic fields
of cosmic objects came with the development of mean--field dynamo theory.
A central component of this theory is the $\alpha$--effect,
that is, a mean electromotive force with a component parallel or antiparallel
to the mean magnetic field in a turbulently moving electrically conducting fluid.
The $\alpha$--effect, which occurs naturally in inhomogeneous turbulence on a rotating body,
is a crucial element of the dynamo mechanisms proposed and widely accepted for the Sun and planets,
for other stellar objects and even for galaxies;
see, e.g., \cite{krauseetal80,ruedigeretal04,brandenburgetal05}.

In the majority of investigations this effect has been merely calculated in the so-called
second--order correlation approximation (SOCA), or first--order smoothing approximation (FOSA),
which ignores all contributions of higher than second order in the turbulent part of the fluid velocity.
Moreover in many cases attention has been focussed on the high--conductivity limit only,
which can be roughly characterized by short correlation times of the turbulent motion in comparison
to the magnetic--field decay time for a turbulent eddy.
Under these circumstances the $\alpha$--effect is closely connected
with some average over the kinetic helicity of the turbulent motion,
that is, of $\bu \cdot \bomega$, where $\bu$ means the turbulent part of the fluid velocity,
$\bomega$ the corresponding vorticity, $\bomega = \bnab \x \bu$;
see, e.g., \cite{krauseetal80,ruedigeretal04,raedler00b,brandenburgetal05,raedleretal07}.
More precisely, the coefficient $\alpha$ for isotropic turbulence turns out
to be equal to $- \frac{1}{3} \int_0^\infty \ol{\bu (\bx, t) \cdot \bomega (\bx, t - \tau)} \, \dd \tau$,
often expressed in the form $- \frac{1}{3} \ol{\bu \cdot \bomega} \, \tau_c$
with equal arguments of $\bu$ and $\bomega$ and some appropriate time $\tau_c$.
Here and in what follows overbars indicate averages.
For anisotropic turbulence the trace of the $\alpha$--tensor is just
equal to three times this value of $\alpha$.

These findings have been sometimes overinterpreted in the sense that mean--field dynamos,
or even dynamos at all, might not work without kinetic helicity of the fluid motion,
that is, if $\bu \cdot \bomega$ vanishes.
There exist however several counter--examples.

Firstly, the mean-field dynamo theory offers also dynamo mechanisms without $\alpha$--effect,
for which the average of the kinetic helicity may well be equal to zero.
For instance, a combination of the so--called $\bOmega \x \bJ$--effect, which may occur even
in the case of homogeneous turbulence in a rotating body, with shear
associated with differential rotation, can act as a dynamo
\cite{raedler69b,raedler70,roberts72,raedler76,stix76b,raedler80,moffattetal82,raedler86,raedleretal03}.
The possibility of a dynamo due to turbulence influenced by large-scale shear and the shear flow
itself \cite{rogachevskiietal03,rogachevskiietal04,rogachevskiietal06b}
is still under debate
\cite{brandenburg05b,raedleretal06,ruedigeretal06,BRRK07}.

Secondly, in dynamo theory beyond the mean--field concept a number of examples of dynamos
without kinetic helicity are known.
The dynamo proposed by Herzenberg \cite{herzenberg58},
usually considered as the first existence proof for homogeneous dynamos at all, works without kinetic helicity.
Likewise in the dynamo models of Gailitis working with an axisymmetric meridional circulation
in cylindrical or spherical geometry \cite{gailitis70,gailitis93,gailitis93b} there is no kinetic helicity.
We mention here further the dynamo in a layer with hexagonal convection cells proposed
by Zheligovsky and Galloway \cite{zheligovskyetal98}, in which the kinetic helicity is zero everywhere.

Thirdly, it is known since the studies by Kazantsev \cite{kazantsev68}
that small--scale dynamos may work without kinetic helicity.
This fact has meanwhile been confirmed by many other investigations;
see, e.g., \cite{brandenburgetal05}.

Let us return to the mean--field concept but
restrict ourselves, for the sake of simplicity, to homogeneous and statistically steady turbulence.
We stay first with the second--order correlation approximation
but relax the restriction to the high--conductivity limit.
As is known from early studies,
e.g. \cite{krauseetal71b,krauseetal80,moffattetal82},
the crucial parameter for the $\alpha$--effect is then no longer
the mean kinetic helicity $\ol{\bu \cdot \bomega}$.
The $\alpha$-effect is in general determined by the function
$h (\bxi, \tau) = \ol{\bu (\bx, t) \cdot \bomega (\bx + \bxi, t + \tau)}$
or, what is equivalent, by the kinetic helicity spectrum,
defined as its Fourier transform with respect to $\bxi$ and $\tau$.
In the high--conductivity limit this brings us back to the above--mentioned results.
In the low conductivity limit,
that is, large correlation times in comparison to the magnetic--field decay time for a turbulent eddy,
the $\alpha$--effect is closely connected with the quantity $\ol{\bu \cdot \bpsi}$,
where $\bpsi$ is the vector potential of $\bu$,
that is $\bnab \x \bpsi = \bu$ with $\bnab \cdot \bpsi = 0$.
For isotropic turbulence the coefficient $\alpha$ is then equal to $ - \frac{1}{3 \eta} \, \ol{\bu \cdot \bpsi}$,
where $\eta$ means the magnetic diffusivity of the fluid
\cite{krauseetal71b,krauseetal80,raedler00b,raedleretal07,suretal07}.
In the anisotropic case the trace of the $\alpha$--tensor is again equal to three times this value of $\alpha$.
In general $\ol{\bu \cdot \bomega}$ does not vanish in this limit but it is without interest
for the $\alpha$--effect.
Both $\ol{\bu \cdot \bomega}$ and $\ol{\bu \cdot \bpsi}$ are quantities, which, if non--zero,
indicate deviations of the turbulence from reflectional symmetry.
In the second--order correlation approximation the $\bOmega \x \bJ$--effect, too,
is completely determined by the kinetic helicity spectrum \cite{moffattetal82}.

Beyond the second--order correlation approximation the $\alpha$--effect is no longer determined
by the kinetic helicity spectrum alone.
In an example given by Gilbert et al.\ \cite{gilbertetal88} an $\alpha$--effect
(with zero trace of the $\alpha$--tensor) occurs in a non-steady flow even with zero kinetic helicity spectrum.
Note that small-scale dynamos may also work with zero kinetic helicity spectrum,
but they do not produce large scale fields.

Interesting simple dynamo models have been proposed
by G.\ O.\ Roberts \cite{robertsgo70,robertsgo72}.
In his second paper \cite{robertsgo72} he considered dynamos due to steady fluid flows
which are periodic in two Cartesian coordinates, say $x$ and $y$, but independent of the third one, $z$.
When speaking of a ``Roberts dynamo" in what follows we refer always
to the first flow pattern envisaged there [equation (5.1), figure 1].
This dynamo played a central role in designing the Karlsruhe dynamo experiment
\cite{muelleretal00,stieglitzetal01,muelleretal02,stieglitzetal02}.
It can be easily interpreted within the mean--field concept.
In that sense it occurs as an $\alpha$--effect dynamo with an anisotropic $\alpha$-effect.
The flow pattern proposed by Roberts, and in a sense realized in the \mbox{Karlsruhe} experiment,
shows non--zero kinetic helicity.
In this paper we want to demonstrate that this kind of dynamo
works also with a slightly modified flow pattern
in which the kinetic helicity is exactly equal to zero everywhere
(but the kinetic helicity spectrum remains non--zero).

In section II we present a simple mean--field theory of Roberts dynamos
at the level of the second--order correlation approximation.
In section III we report on numerical results that apply independently of this approximation.
Finally, in section IV some conclusions are discussed.

\section{A modified Roberts dynamo}
\label{sec2}

\subsection{Starting point}

\begin{figure}
\includegraphics[width=.45\textwidth]{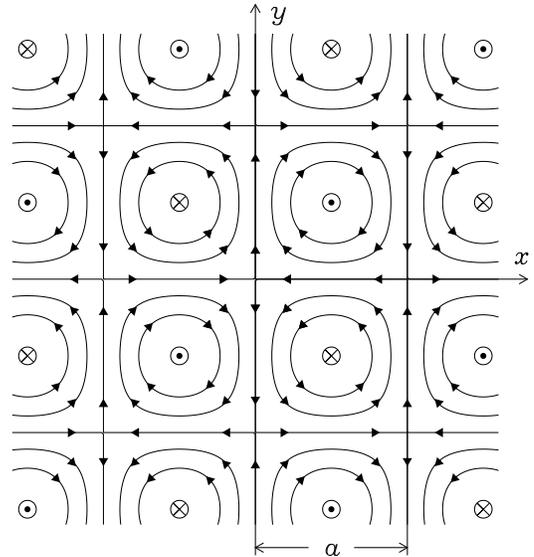}
\caption[]
{The Roberts flow pattern}
\label{robflow}
\end{figure}

We focus our attention now on the Roberts dynamo in the above sense
and refer again to a Cartesian coordinate system $(x, y, z)$.
The fluid is considered as incompressible.
Therefore its velocity, $\bu$, has to satisfy $\bnab \cdot \bu = 0$, and it
is assumed to be a sum of two parts, one proportional to $\be \x \bnab \chi$
and the other to $\chi \be$, where $\be$ is the unit vector in $z$ direction,
and $\chi = \sin (\pi x / a) \sin (\pi y / a)$.
The corresponding flow pattern is depicted in figure \ref{robflow}.
We denote the sections defined by $n a \leq x \leq (n+1) a$ and $m a \leq x \leq (m+1) a$
with integer $n$ and $m$ as ``cells".
In that sense the velocity $\bu$ changes sign when we proceed
from one cell to an adjacent one.
A fluid flow of that kind acts indeed as a dynamo.
Magnetic field modes with the same periodicity in $x$ and $y$
varying however with $z$ like $\sin (k z)$ or $\cos (k z)$
may grow for sufficiently small values of $|k|$.
The most easily excitable modes for a given $k$ possess a part
which is independent of $x$ and $y$, but varies with $z$.

\subsection{Mean--field theory with a generalized fluid flow}

Let us proceed to more general flow patterns which show some of the crucial properties
of the specific flow pattern considered so far.
We assume again that the flow is incompressible, steady,
depends periodically on $x$ and $y$ but is independent of $z$,
and that it shows a cell structure as in the example above.
More precisely we require that $\bu$ changes sign if the pattern is shifted by a length $a$
along the $x$ or $y$ axes or is rotated by $90^\circ$ about the $z$ axis.

Let us sketch the mean--field theory of dynamos working with fluid flows of this type
(partially using ideas described in our paper \cite{raedleretal02},
in the following referred to as RB03).

It is assumed that the magnetic field $\bB$ is governed by the induction equation
\begin{equation}
\eta \bnab^2 \bB + \bnab \x (\bu \x \bB)- \p_t \bB = \bzo \, , \quad \bnab \cdot \bB = 0,
\label{eq01}
\end{equation}
with the magnetic diffusivity $\eta$ being constant.

We define mean fields by averaging the original fields over a square
that corresponds to four cells in the $xy$ plane.
More precisely we define the mean field $\ol{F}$, which belongs to an original field $F$,
in any given point $\bx = \bx_0$ by averaging $F$ over the square defined
by $x_0 - a \leq x \leq x_0 + a$
and $y_0 - a \leq y \leq y_0 + a$.
If applied to quantities which are periodic in $x$ and $y$ with a period length $2a$ this average
satisfies the Reynolds rules.
Clearly we have $\ol{\bu} = \bzo$.

When taking the average of equation (\ref{eq01}) and denoting $\bB - \bmB$ by $\bb$
we obtain the mean--field induction equation
\begin{equation}
\eta \bnab^2 \bmB + \bnab \x \bscE - \p_t \bmB = \bzo \, , \quad \bnab \cdot \bmB = 0,
\label{eq03}
\end{equation}
with the mean electromotive force
\begin{equation}
\bscE = \ol{\bu \x \bb} \, .
\label{eq05}
\end{equation}
For the determination of $\bscE$ for a given $\bu$ the equation for $\bb$ is of interest.
This equation follows from (\ref{eq01}) and (\ref{eq03}),
\begin{eqnarray}
\eta \bnab^2 \bb + \bnab \x (\bu \x \bb)' - \p_t \bb &=& - \bnab \x (\bu \x \bmB) \, ,
\nonumber\\
\bnab \cdot \bb &=& 0 \, ,
\label{eq07}
\end{eqnarray}
where $(\bu \x \bb)'$ stands for $\bu \x \bb - \ol{\bu \x \bb}$.

For the sake of simplicity we consider here only the steady case.
We introduce a Fourier transformation with respect to $z$,
that is, represent any function $F (x,y,z)$
by $F (x,y,z) = \int \hat{F} (x,y, k) \exp (\iu k z) \, \dd k$.
If $F (x,y,z)$ is real, we have $\hat{F}^* (x,y,k) = \hat{F} (x,y,-k)$,
where the asterisk means complex conjugation.
We have now
\begin{equation}
\hat{\bscE} = \ol{\bu \x \hat{\bb}} \, .
\label{eq08}
\end{equation}
Transforming (\ref{eq07}) accordingly we obtain
\begin{eqnarray}
\eta (\bnab^2 - k^2) \hat{\bb} + (\bnab + \iu k \be) \x (\bu \x \hat{\bb})' &&
\nonumber\\
&& \!\!\!\!\!\!\!\!\!\!\!\!\!\!\!\!\!\!\!\!\!\!\!\!\!\!
     \!\!\!\!\!\!\!\!\!\!\!\!\!\!\!\!\!\!\!\!\!\!\!\!\!
     = - (\bnab + \iu k \be) \x (\bu \x \hat{\bmB}) \, ,
\nonumber\\
\bnab \cdot \hat{\bb} + \iu k \hat{b}_z &=& 0 \, .
\label{eq09}
\end{eqnarray}
We further suppose that $\hat{\bmB}$ is independent of $x$ and $y$.
This implies that $\hat{\bscE}$ is independent of $x$ and $y$, too.
It is then obvious that the connection between $\hat{\bscE}$ and $\hat{\bmB}$ must have the form
\begin{equation}
\hat{\cE}_i = \hat{a}_{ij} \hat{\mB}_j \, ,
\label{eq11}
\end{equation}
where $\hat{a}_{ij}$ is, like $\hat{\bscE}$ and $\hat{\bmB}$, independent
of $x$ and $y$, but depends on $k$.
Because $\hat{\cE}_i^* (k) = \hat{\cE}_i (-k)$ and $\hat{\mB}_i^* (k) = \hat{\mB}_i (-k)$ it has also to satisfy
$\hat{a}_{ij}^* (k) = \hat{a}_{ij} (- k)$.

Clearly $\hat{a}_{ij}$ is determined by $\bu$.
Since $\bscE$ does not change under inverting the sign of $\bu$,
and a $90^\circ$ rotation of the $\bu$ field about the $z$ axis changes nothing else than its sign,
the components of the tensor $\hat{a}_{ij}$ have to be invariant under $90^\circ$ rotations of the coordinate system
about the $z$ axis.
We may therefore conclude that
\begin{equation}
\hat{a}_{ij} = a_1 (|k|) \, \delta_{ij} + a_2 (|k|) \, e_i e_j + \iu \, a_3 (|k|) \, k \, \epsilon_{ijk} e_k
\label{eq12}
\end{equation}
with real $a_1$, $a_2$ and $a_3$.
Together with (\ref{eq11}) this leads to $\hat{\cE}_z = (a_1 + a_2) \hat{\mB}_z$.
On the other hand, $\hat{\cE}_z$ is equal to the average of $u_x \hat{b}_y - u_y \hat{b}_x$,
and we may conclude from (\ref{eq09}) that $\hat{b}_x$ and $\hat{b}_y$ are independent of $\hat{\mB}_z$.
This implies $a_1 + a_2 = 0$.
We may then write
\begin{equation}
\hat{a}_{ij} = - \hat{\alpha}_\perp (k) \, (\delta_{ij} - e_i e_j)
     + \iu \, \hat{\beta} (k) k \epsilon_{ijk} e_k \, ,
\label{eq13}
\end{equation}
with two real quantities $\hat{\alpha}_\perp$ and $\hat{\beta}$,
which are even functions of $k$.
From (\ref{eq11}) and (\ref{eq13}) we obtain
\begin{equation}
\hat{\bscE} = - \hat{\alpha}_\perp (k) \big(\hat{\bmB} - (\be \cdot \hat{\bmB}) \, \be \big)
      - \iu \, \hat{\beta} \, k  \be \x \hat{\bmB} \, .
\label{eq15}
\end{equation}
This relation is discussed in some detail in RB03, section V B.

We now restrict our attention to the limit of small $k$, that is, small variations of $\bmB$ with $z$.
Then $\hat{\alpha}_\perp$ and $\hat{\beta}$ lose their dependence on $k$.
Denoting them by $\alpha_\perp$ and $\beta$ we conclude from (\ref{eq15}) that
\begin{equation}
\bscE = - \alpha_\perp \big(\bmB - (\be \cdot \bmB) \, \be \big) - \beta \, \be \x \dd \bmB / \dd z \, .
\label{eq17}
\end{equation}

Equation (\ref{eq03}) together with (\ref{eq17}) allow solutions of the form
$\bmB = \Re [\bmB_0 \exp(\iu k z + \lambda t)]$
with $\lambda = \pm \alpha_\perp k - (\eta + \beta) k^2$.
Growing solutions are possible if
\begin{equation}
\frac{|\alpha_\perp|}{|k| (\eta + \beta)} > 1 \, .
\label{eq19}
\end{equation}
We note that, in agreement with the anti--dynamo theorem by Zeldovich, $\lambda$ vanishes
if $k \to 0$,
that is, if $\bmB$ loses its dependence on $z$.

\subsection{Calculation of $\alpha_\perp$ and $\beta$}

For the determination of the coefficients $\alpha_\perp$ and $\beta$ we restrict ourselves
to an approximation which corresponds to the second--order correlation approximation.
It is defined by the neglect of $(\bu \x \bb)'$ in (\ref{eq07}), or $(\bu \x \hat{\bb})'$ in (\ref{eq09}).
In view of $\alpha_\perp$ and $\beta$, which correspond to the limit of small $k$,
we expand $\hat{\bb}$ and $\hat{\bmB}$ in powers of $k$
but neglect all contributions with higher than first powers of $k$.
That is, $\hat{\bb} = \hat{\bb}^{(0)} +  k \hat{\bb}^{(1)}$
and $\hat{\bmB} = \hat{\bmB}^{(0)} +  k \hat{\bmB}^{(1)}$,
where of course $\hat{\bb}^{(0)} = \bb^{(0)}$ and $\hat{\bmB}^{(0)} = \bmB^{(0)}$.
From (\ref{eq09}) follows then
\begin{eqnarray}
\eta \bnab^2 \hat{\bb}^{(0)} &=& - (\hat{\bmB}^{(0)} \cdot \bnab) \, \bu
\nonumber\\
\eta \bnab^2 \hat{\bb}^{(1)} &=& - \iu \, \be \x (\bu \x \hat{\bmB}^{(0)})
     - (\hat{\bmB}^{(1)} \cdot \bnab) \, \bu \, .
\label{eq21}
\end{eqnarray}
We put now $\bu = \bnab \x \bpsi$, where $\bpsi$ denotes a vector potential satisfying $\bnab \cdot \bpsi = 0$,
further $\bpsi = \bnab \x \tilde{\bpsi}$ with $\bnab \cdot \tilde{\bpsi} = 0$,
which implies
\begin{equation}
\bu = - \bnab^2 \tilde{\bpsi} \, .
\label{eq23}
\end{equation}
From (\ref{eq21}) with $\bu$ expressed in this way we conclude
\begin{eqnarray}
\hat{\bb}^{(0)} &=& \frac{1}{\eta} \, (\hat{\bmB}^{(0)} \cdot \bnab) \, \tilde{\bpsi}
\nonumber\\
\hat{\bb}^{(1)} &=&  \frac{1}{\eta} \, \Big[ \iu \, \be \x (\tilde{\bpsi} \x \hat{\bmB}^{(0)})
     - (\hat{\bmB}^{(1)} \cdot \bnab) \tilde{\bpsi} \Big] \, .
\label{eq25}
\end{eqnarray}
Putting in the above sense also $\hat{\bscE} = \hat{\bscE}^{(0)} + k \hat{\bscE}^{(1)}$ we find
\begin{eqnarray}
\hat{\bscE}^{(0)} &=& \frac{1}{\eta} \, \ol{\bu \x (\hat{\bmB}^{(0)} \cdot \bnab) \, \tilde{\bpsi}}
\nonumber\\
\hat{\bscE}^{(1)} &=&  - \frac{1}{\eta} \,
     \Big\{ \iu \, \be \x \Big[\ol{\bu \x (\tilde{\bpsi} \x \hat{\bmB}^{(0)}}) \Big]
\label{eq27}\\
     && \qquad + \ol{\bu \x (\hat{\bmB}^{(1)} \cdot \bnab) \tilde{\bpsi}} \, \Big\} \, .
\nonumber
\end{eqnarray}

We compare this with what follows from (\ref{eq15}) in this expansion,
\begin{eqnarray}
\hat{\bscE}^{(0)} &=& - \alpha_\perp \Big[\hat{\bmB}^{(0)} - (\be \cdot \hat{\bmB}^{(0)}) \, \be\Big]
\nonumber\\
\hat{\bscE}^{(1)} &=& - \iu \, \beta \, \be \x \hat{\bmB}^{(0)}
     - \alpha_\perp \Big[\hat{\bmB}^{(1)} - (\be \cdot \hat{\bmB}^{(1)}) \, \be\Big] \, . \,
\label{eq29}
\end{eqnarray}
In this way we obtain first
\begin{eqnarray}
\alpha_\perp &=& - \frac{1}{\eta} \, \big(\ol{\bu \x (\bg \cdot \bnab) \tilde{\bpsi}} \big) \cdot \bg
\nonumber\\
\beta &=& - \frac{1}{\eta} \, \Big[\, \ol{\bu \cdot \tilde{\bpsi}}
- \ol{(\bh\cdot \bu) \, (\bh \cdot \tilde{\bpsi})} \,\Big] \, ,
\label{eq31}
\end{eqnarray}
where $\bg$ and $\bh$ are unit vectors in the directions
of $\hat{\bmB}^{(0)} - (\be \cdot{\hat{\bmB}}^{(0)}) \, \be$
and $\be \x \hat{\bmB}^{(0)}$, respectively.
Of course, $\alpha_\perp$ and $\beta$ cannot really depend on these directions,
and we may average the two expressions for $\alpha_\perp$ with $\bg = (1, 0, 0)$
and $\bg = (0, 1, 0)$
and likewise those for $\beta$ with $\bh = (1, 0, 0)$ and $\bh = (0, 1, 0)$.
This yields
\begin{eqnarray}
\alpha_\perp &=& \frac{1}{2 \eta} \, \ol{\bu \cdot (\bnab \x \tilde{\bpsi})},
\nonumber\\
\beta &=& - \frac{1}{2 \eta} \, \big( \ol{\bu \cdot \tilde{\bpsi}}
     + \ol{(\be \cdot \bu) \, (\be \cdot \tilde{\bpsi})} \big) \, .
\label{eq33}
\end{eqnarray}
The result for $\alpha_\perp$ can also be written in the form
\begin{equation}
\alpha_\perp  =  \frac{1}{2 \eta} \, \ol{\bu \cdot \bpsi}
    = \frac{1}{2 \eta} \, \ol{\bpsi \cdot (\bnab \x \bpsi)} \, .
\label{eq35}
\end{equation}
In agreement with our remarks in the introduction the $\alpha$--effect is not determined
by the average of the kinetic helicity $\ol{\bu \cdot \bomega}$
but by the related but different quantity $\ol{\bu \cdot \bpsi}$,
that is, some kind of mean helicity of the vector potential $\bpsi$.

\subsection{A specific flow}

\begin{figure}
\includegraphics[width=.3\textwidth]{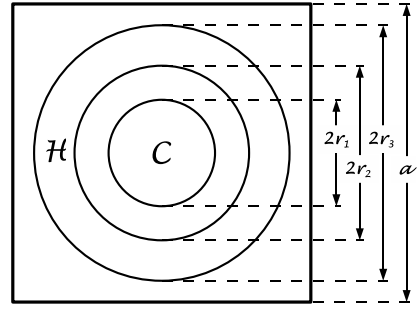}
\caption[]
{A single cell of the modified Roberts flow}
\label{cell}
\end{figure}

We elaborate now on the above result for $\alpha_\perp$ for flow patterns
similar to those which have been considered in the context of the Karlsruhe dynamo experiment
\cite{raedleretal02b}.
To define the flow first in a single cell like those
depicted in figure \ref{cell} we introduce a cylindrical
coordinate system $(r, \varphi, z)$ with $r = 0$ in the middle of this cell and $z$ as in the
Cartesian system used above.
We put then
\begin{eqnarray}
u_r &=& 0 \qquad \qquad \qquad \quad \;\;\;\; \mbox{everywhere}
\nonumber\\
u_\varphi &=& 0 \, , \;\; u_z = u \qquad \quad \;\;\;\; \mbox{in} \;\; r \leq r_1
\nonumber\\
u_\varphi &=& 0 \, , \;\; u_z = 0 \qquad \qquad \; \mbox{in} \;\; r_1 < r < r_2
\label{eq41}\\
u_\varphi &=& \omega r \, , \;\; u_z = \varepsilon \omega a / 2 \quad \;\;
    \mbox{in} \;\, r_2 \leq r \leq r_3
\nonumber
\end{eqnarray}
with $\omega$ and $\varepsilon$ being constants.
The full flow pattern is defined by continuation of that in the considered cell
to the other cells such that the flow has always different signs in two adjacent cells.
Thinking of the ``central channels" and the ``helical channels" of the Karlsruhe device we label
the flow regions $r \leq r_1$  and $r_2 \leq r \leq r_3$ by $\cal{C}$ and $\cal{H}$,
respectively.
Clearly finite kinetic helicity exists only in $\cal{H}$, and only for non--zero $\varepsilon$.
Note that, in contrast to the situation depicted in figure~\ref{robflow}, here the kinetic helicity is non--negative
as long as $\varepsilon$ is positive.

For the flow defined in this way quantities like $\ol{\bu \cdot \bomega}$
or $\ol{\bu \cdot \bpsi}$ depend on $u$ and $\omega$.
In the following we express $u$ by the flow rate $V_\mathrm{C}$ of fluid
through the cross section $0 \leq r \leq r_1$ at a given $z$,
and $\omega$ by the flow rate $V_\mathrm{H}$ the fluid through a meridional surface
defined by $r_2 \leq r \leq r_3$ and $0 \leq z \leq a$ and a given $\varphi$.
As can be easily verified we have then
\begin{equation}
\ol{\bu \cdot \bomega} = c \, \varepsilon \, V_\mathrm{H}^2 \, ,
\label{eq45}
\end{equation}
with some positive quantity $c$ depending on $r_2$, $r_3$ and $a$.
For $\ol{\bu \cdot \bpsi}$ it is important that the part of the vector potential $\bpsi$
resulting from the flow in the region $\cal{C}$ of a given cell does not vanish in $\cal{H}$,
and likewise the part resulting from the flow in $\cal{H}$ does not vanish in $\cal{C}$.
Moreover, the parts of $\bpsi$ resulting from flows in other cells have to be taken into account.
Considering these aspects we may conclude that
\begin{equation}
\ol{\bu \cdot \bpsi} = (c_1 V_\mathrm{C} + \varepsilon c_2 V_\mathrm{H}) \, V_\mathrm{H}
\label{eq47}
\end{equation}
with non--zero quantities $c_1$ and $c_2$ depending on $r_1$, $r_2$, $r_3$ and $a$.

Clearly, $\ol{\bu \cdot \bpsi}$ and, according to (\ref{eq35}), $\alpha_\perp$
remain in general different from zero if the flow loses its kinetic helicity,
that is, as $\ol{\bu \cdot \bomega}$, or $\varepsilon$, vanish.
This implies the possibility of $\alpha$--effect dynamos without kinetic helicity.

\section{Numerical examples of $\alpha$--effect dynamos with and without kinetic helicity}
\label{sec3}

In our above calculations of $\alpha_\perp$ and $\beta$ an approximation
in the spirit of the second--order correlation approximation has been used.
The results are reliable only if a suitably defined magnetic Reynolds number
is much smaller than unity.
This can also be expressed by requiring that the normalized flow rates
$\tilde{V}_\mathrm{C} = V_\mathrm{C} / \eta a$
and $\tilde{V}_\mathrm{H} = V_\mathrm{H} / \eta a$ show this property.
In addition we have assumed weak variations of the mean magnetic field in the $z$--direction,
more precisely $\kappa = k/a \ll 1$.

In order to confirm the existence of $\alpha$--effect dynamos without kinetic helicity
in an independent way and to check whether it occurs only under the mentioned conditions
or over a wider range of parameters,
equation (\ref{eq01}) with the flow defined by (\ref{eq41}) has been solved numerically.
The same numerical method as in RB03 has been used.

\begin{table}
\caption{Examples of marginal dynamo states with $\kappa = 0.9$
\label{tab1}}
\begin{tabular}{ccccccc}
\hline
case & ~$2r_1/a$~ & ~$2r_2/a$~ & ~$2r_3/a$~ & $\varepsilon$ & $\tilde{V}_\mathrm{C}$ & $\tilde{V}_\mathrm{H} $ \cr
\hline
  (i) & 0.5 & 0.5 & 1.0 & 0.228 & 2 & 0.736  \cr
 (ii) & 0.5 & 0.5 & 1.0 & 0 & 2 & 0.805  \cr
(iii) & 0.5 & 0.7 & 1.0 & 0 &  2 & 0.965 \cr
\hline
\end{tabular}
\end{table}

\begin{figure}
\includegraphics[width=.5\textwidth]{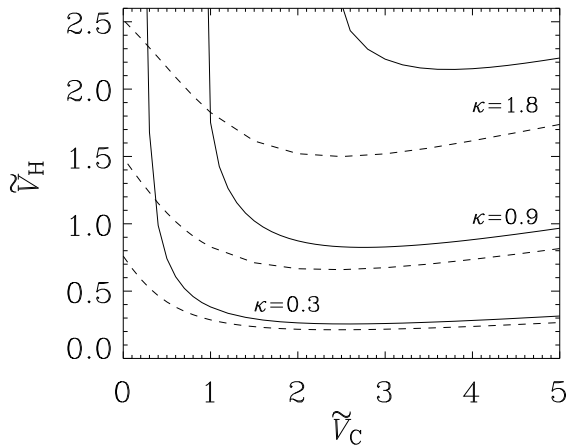}\caption[]
{Lines of marginal stability in the $\tilde{V}_\mathrm{C} \, \tilde{V}_\mathrm{H}$ plane
for the cases (i) (dashed lines) and (iii) (solid lines) defined in Table~\ref{tab1}
and three values of $\kappa$.}
\label{ptable_crit}
\end{figure}

In table~\ref{tab1} and figure~\ref{ptable_crit} some numerically determined dimensionless flow rates
$\tilde{V}_\mathrm{C}$ and $\tilde{V}_\mathrm{H}$ are given for marginal dynamo states.
For case (i) a non-zero kinetic helicity was chosen, for the cases (ii) and (iii) zero kinetic helicity.
The marginal modes turned out to be non--oscillatory.
Note that the parameters for case (i),
$2 r_1 / a = 2 r_2 / a = 0.5$, $2 r_3 / a = 1$, $\varepsilon = 0.228$ and $\kappa = 0.9$,
are realistic values in view of the Karlsruhe dynamo device.
In the cases (i) and (ii) the regions $\cal{C}$ and $\cal{H}$ have a common surface.
In order to make sure that there is no effect of an artificial kinetic helicity
due to the restricted numerical resolution at this surface, in case (iii) a clear separation
of these two regions was considered.

As long as the kinetic helicity in $\cal{H}$ does not vanish, that is $\varepsilon \not= 0$,
the dynamo works even with $V_\mathrm{C} = 0$, that is zero flow in $\cal{C}$.
In the absence of kinetic helicity, $\varepsilon = 0$, the dynamos works, too, as long as $V_\mathrm{C}$
and therefore $\ol{\bu \cdot \bpsi}$ are different from zero.
However, the dynamo disappears if, in addition to $\varepsilon$, also $V_\mathrm{C}$
and so $\ol{\bu \cdot \bpsi}$ vanish.
This is plausible since then choices of the coordinate origin $\bx = \bzo$ are possible
such that $\bu (\bx)$ and $-\bu (-\bx)$ coincide.
In this case, usually referred to as ``parity--invariant", any $\alpha$--effect can be excluded.

The dissipation of the mean magnetic field grows with $\kappa$.
By this reason the dynamo threshold grows with $\kappa$, too.

\section{Conclusions}

We have examined a modified version of the Roberts dynamo
with a steady fluid flow as it was considered
in the context of the \mbox{Karlsruhe} dynamo experiment.
By contrast to what might be suggested by simple explanations of this dynamo the necessary deviation
of the flow from reflectional symmetry is not adequately described
by the kinetic helicity $\bu \cdot \bomega$ of the fluid flow or its average $\ol{\bu \cdot \bomega}$.
Dynamo action with steady fluid flow
requires in truth a deviation which is indicated by
a non--zero value of the quantity $\ol{\bu \cdot \bpsi}$,
where $\bpsi$ is the vector potential of $\bu$.
Even if the kinetic helicity is equal to zero everywhere, dynamo action occurs
if only the quantity $\ol{\bu \cdot \bpsi}$, the helicity of the vector potential of the flow,
is unequal to zero.
This fact is not surprising in light of the general findings of mean--field electrodynamics.

We stress that our result does not mean that the quantity $\ol{\bu \cdot \bpsi}$ is in general
more fundamental for the $\alpha$--effect and its dynamo action than $\ol{\bu \cdot \bomega}$.
In cases with steady fluid flow indeed $\ol{\bu \cdot \bpsi}$ is crucial.
However, with flows varying rapidly in time, the crucial quantity is $\ol{\bu \cdot \bomega}$.

Both the mean kinetic helicity $\ol{\bu \cdot \bomega}$ and the quantity $\ol{\bu \cdot \bpsi}$
are determined by the helicity spectrum of the flow.
That is, the Roberts dynamo does not work with zero kinetic helicity spectrum.
In this sense the situation is different from that in the interesting example
of an $\alpha$--effect dynamo
given by Gilbert et al.\ \cite{gilbertetal88}, which works with a very special
non--steady flow for which the kinetic helicity spectrum is indeed zero.

Looking back at the Karlsruhe experiment we may state that a modified
(technically perhaps more complex) version without kinetic helicity of the fluid flow
should show the dynamo effect, too.
As table~\ref{tab1} exemplifies, however, the kinetic helicity can well reduce the dynamo threshold.
In that sense we may modify the title of the paper of Gilbert et al.\ and say:
Helicity is unnecessary {\it for Roberts dynamos} -- but it helps!

\bigskip

{\bf Acknowledgement}
The results reported in this paper have been obtained during stays of KHR at NORDITA.
He is grateful for its hospitality.
AB thanks the Kavli Institute for Theoretical Physics for hospitality.
This research was supported in part by the National Science
Foundation under Grant No.\ PHY05-51164.

\end{document}